\documentclass[10pt,twocolumn,letterpaper]{article}

\usepackage{iccv}
\usepackage{times}
\usepackage{epsfig}
\usepackage{graphicx}
\usepackage{amsmath}
\usepackage{amssymb}
\usepackage{slashbox}
\bgroup

\usepackage{hyperref}

\iccvfinalcopy 

\def\iccvPaperID{7} 

\ificcvfinal\fi
\begin{document}
\def\Ali#1{{\color{blue}{\bf [Ali:} {\it{#1}}{\bf ]}}}
\def\LP#1{{\color{red}{\bf [LP:} {\it{#1}}{\bf ]}}}

\title{Super-resolved Chromatic Mapping of Snapshot Mosaic Image Sensors via a Texture Sensitive Residual Network}

\author{Mehrdad Shoeiby\\
DATA61-CSIRO\\
ACT, Australia\\
{\tt\small mehrdad.shoeiby@data61.csiro.au}
\and
Lars Petersson\\
Data61-CSIRO\\
ACT, Australia\\
{\tt\small lars.petersson@data61.csiro.au}
\and
Mohammad Ali Armin\\
Data61-CSIRO\\
ACT, Australia\\
{\tt\small ali.armin@data61.csiro.au}
\and
Sadegh Aliakbarian\\
Data61-CSIRO\\
ACT, Australia\\
{\tt\small sadegh.aliakbarian@data61.csiro.au}
\and
Antonio Robles-Kelly\\
Deakin University\\
VIC, Australia\\
{\tt\small antonio.robbles-kelly@deakin.edu.au}
}
\maketitle

\begin{abstract}
This paper introduces a novel method to simultaneously super-resolve and colour-predict images acquired by snapshot mosaic sensors. These sensors allow for spectral images to be acquired using low-power, small form factor, solid-state CMOS sensors that can operate at video frame rates without the need for complex optical setups. Despite their desirable traits, their main drawback stems from the fact that the spatial resolution of the imagery acquired by these sensors is low. Moreover, chromatic mapping in snapshot mosaic sensors is not straightforward since the bands delivered by the sensor tend to be narrow and unevenly distributed across the range in which they operate. We tackle this drawback as applied to chromatic mapping by using a residual channel attention network equipped with a texture sensitive block. Our method significantly outperforms the traditional approach of interpolating the image and, afterwards, applying a colour matching function. This work establishes  state-of-the-art in this domain while also making available to the research community a dataset containing 296 registered stereo multi-spectral/RGB images pairs. 
\end{abstract}

\section{Introduction}
\begin{figure}[t]
  \includegraphics[width=\linewidth]{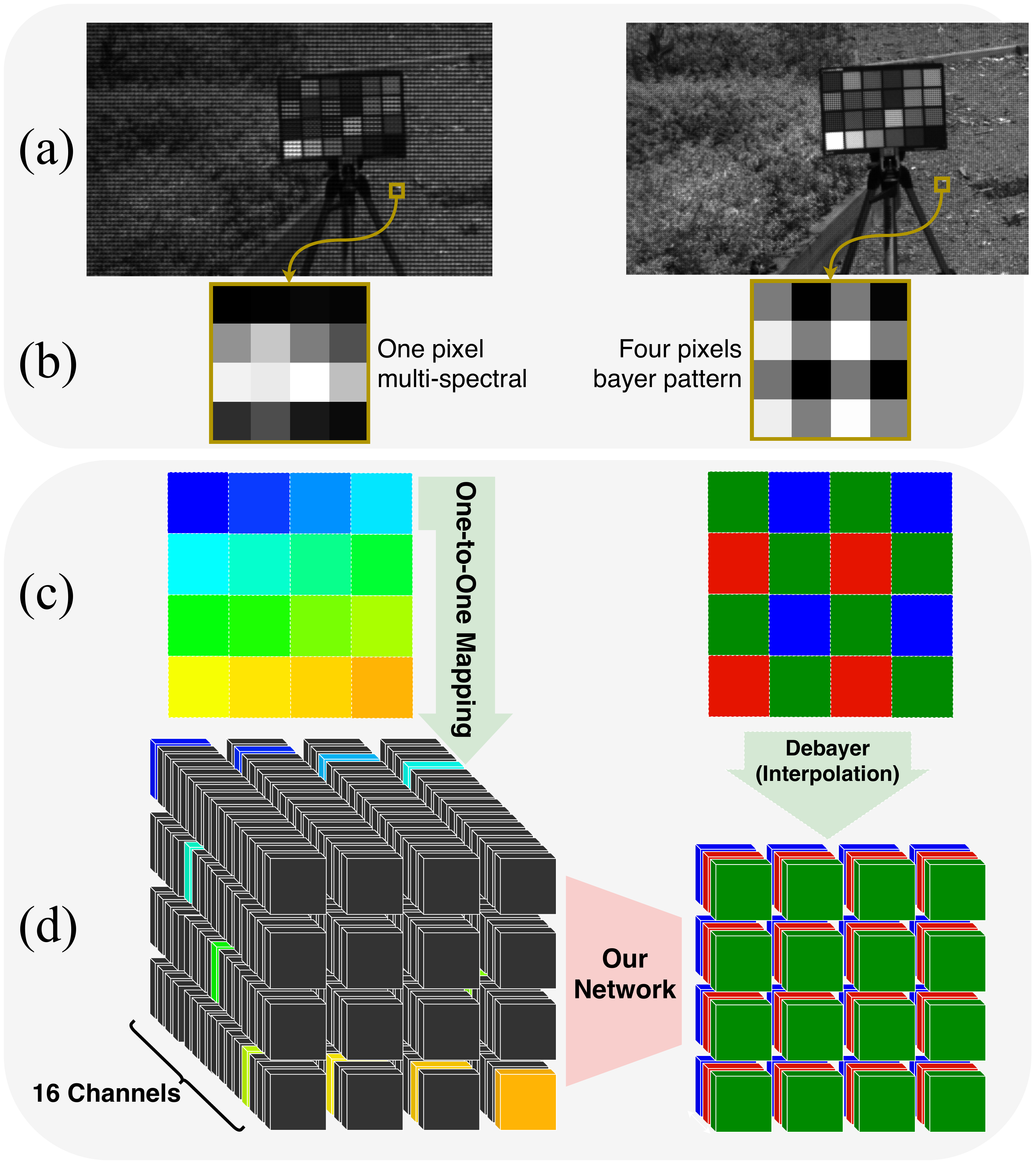}
  \caption{(a) Original mosaic images, with (b) the actual one pixel of the multi-spectral vs 4 pixels of the RGB bayer image encompassing 16 vs 3 channels (wavelengths) per pixel respectively. (c) and (d) demonstrate the data formation for multi-spectral vs RGB images. Each colour is indicative of respective wavelength. The multi-spectral cube is formed by a one-to-one mapping of each wavelength (sub-pixel) to respective channel (one of 16) and zero padding the other 15 sub-pixels. The RGB image is debayered (interpolated) to form the respective R-G-B channels. While a one-to-one mapping (with zero-padding) leads to a large number of redundant zero pixels, as opposed to debayering for RGB images, it results in better super-resolution by taking into account the spatial offset of each pixel.}
  \label{fig:16color}
\end{figure}  

Imaging spectroscopy devices can capture an information-rich representation of a scene, often in terms of tens or hundreds of wavelength-indexed bands. Recent advances in imaging spectroscopy have seen the development of real-time snapshot mosaic image sensors, which are compact in size and exhibit comparable frame rates to current trichromatic cameras ~\cite{cdd_hyper_2013,cmos_review_2006}. Despite of the extensive interest in snapshot mosaic sensors and their potential for multi-spectral imaging, they suffer from an inherent trade-off between the spatial and spectral resolution.  This is as a result of their architecture, where the raw resolution of the detector is distributed across the number of wavelength-indexed bands in the spectral image produced at output. 

As a result,  a higher spatial resolution (smaller pixel size) reduces the number of wavelength bins that can fit on that pixel on the image sensor. This creates a constraint for certain applications where smaller/lighter cameras are needed, for instance, on a UAV~\cite{doering_2016}. A smaller/lighter camera for portability reasons renders a device suffering from lower spatial and spectral resolution. Further, these sensors have promising applications ranging from remote sensing~\cite{msi_remote_2009,msi_remote_2010} to food monitoring~\cite{msi_food_2012}, and from astronomy~\cite{msi_astronomy_2016} to object detection in autonomous vehicles \cite{takumi_2017,lars_2013}.   

Furthermore, in many applications it is useful, or in fact crucial, to obtain an RGB image of the same scene. There is a large body of work in computer vision that can be directly leveraged if we can devise a method that delivers a high-quality, high-resolution RGB image from a multi-spectral sensor. For example, in case of object detection in autonomous vehicles~\cite{doering_2016,lars_2013}, RGB cameras as well as multi-spectral cameras were deployed. The acquired RGB images are usually registered against their multi-spectral counterpart to obtain 3D information of the scene, or to compensate for lower spectral information of the multi-spectral images. However, the low spatial/spectral resolution of multi-spectral cameras could render the registration challenging. 


Traditionally, the RGB equivalent of a scene can be extracted from the spectral image using a Colour Matching Function~\cite{smith_cie_1931}, given that the wavelength range of the camera covers, relatively densely, the complete range of the visual spectrum. Given the relatively limited spectral resolution of snapshot mosaic sensors and their uneven spacing over the operating spectral range, the problem of obtaining high resolution RGB images from low spatial and spectral resolution multi-spectral images is an interesting one that needs to be addressed. Therefore, in this paper, we identify a gap in the scientific literature and propose a single unifying method that carries out 1) color-prediction, and 2) super-resolution (SR) from the multi-spectral space to the RGB space simultaneously. 

The reason for this gap in the literature is the fact that these devices, with relatively recent technology, have just come in to the market. This leads to a limited exposure to researchers and a lack of rich, publicly available, datasets. Thus, we not only present a method that can simultaneously super-resolve and colour-predict spectral images acquired by snapshot mosaic sensors, but also  introduce a novel stereo registered multi-spectral/RGB dataset. Further, our method is quite general in nature, being applicable not only to mosaic snapshot sensor imagery but also to spectral images delivered by other kinds of cameras.


\subsection*{Contributions}
 \begin{itemize}
 \setlength\itemsep{-0.1em}
     \item We propose a method which exploits the mosaic structure of the images acquired by the snapshot sensor directly as opposed to demosaicing images to perform SR and colour prediction sequentially.
     \item We introduce a novel algorithm, to our knowledge the first in the literature, to carry out SR and colour-prediction simultaneously from mosaic images, establishing state-of-the-art in the field.
     \item We introduce a novel dataset containing 296 registered stereo snapshot mosaic-RGB image pairs.
 \end{itemize}

\section{Related work}

Due to the lack of suitable datasets, color-prediction is not a problem that has been extensively studied for multi-spectral images in general. Most of the work is carried out with simulated multi-spectral images \cite{monno2015practical,monno2015n,miao2006binary,jaiswal2016adaptive}. The images were simulated exploiting high resolution hyperspectral images. In addition, the focus of these works, while producing RGB images from simulated multi-spectral images, is to mitigate the structural artifacts introduced by different demosaicing methods. They achieve this, with demonstrated good results, via using forms of interpolation (eg, linear, polynomial, low pass filtering in the frequency domain) to insert additional pixels between the observed spatial/spectral ones. None of the works above attempt to predict RGB from spectrally under-sampled data. The camera we are using is an off-the-shelf commercial camera with narrow FWHM $\approx 15nm$. In addition, it covers the blue and red spectra only partially (see Figure \ref{fig:camera_channels}).  Furthermore, many demosaicing methods such as \cite{monno2015practical,monno2015n} are dependent on a given mosaic pattern as part of their approach. Also,  \cite{monno2015practical,monno2015n,miao2006binary,jaiswal2016adaptive} and most of the demosaicing algorithms rely on a wavelength channel being more densely sampled than the others, using that as a guide image. A $4\times 4$ pattern, with each of the 16 pixels/wavelengths appearing only once (similar to our camera), would be considered by \cite{jaiswal2016adaptive} as severely under-sampled and as demonstrated experimentally, leads to poor results \cite{jaiswal2016adaptive}. Note that a multi-spectral camera that covers a broader spectral range by using a larger number of narrow wavelength bins would be rendered very bulky and expensive, while the camera used here\footnote{Ximea model MQ022HG-IM-SM4x4 470-620} has dimensions of $26mm\times 26mm\times 26mm$.

Image super-resolution (SR) is a problem that has been studied extensively for RGB images. While these algorithms are not perfect for multi-spectral images, they could be exploited to design efficient multi-spectral SR methods. Early approaches to SR were often based upon the rationale that images with higher spatial information have a frequency domain response whose higher frequency components contribute more compared to images with lower spatial information. Hence, such methods~\cite{tsai:84} utilise the shift and aliasing properties of the Fourier transform to obtain a high-resolution representation of the image. Kim~\etal~\cite{kim:90} further extended the concept in~\cite{tsai:84} to take into account noise and spatial blurring present in the input image. In a related development, in~\cite{bose:93}, Tikhonov regularization was exploited to carry out SR in the frequency domain.

Modern single-image methods, often based upon learning, also known as example-based single image SR aim at learning the relationship between low resolution (LR) and high resolution (HR) images by training with LR and HR image pairs. Dong~\etal~\cite{dong2016image} present a deep convolutional network for single-image SR which surpasses the state-of-the-art performance at that time represented by patch-based methods using sparse coding~\cite{yang2010image} or anchored neighborhood regression~\cite{radu_dataset_2014}. Kim~\etal~\cite{kim_vdsr_2016} go deeper with a network based on VGG-net~\cite{simonyan:2014}. The network in~\cite{kim_vdsr_2016} is comprised of 20 layers so as to exploit the image context across larger image regions. More recently, thanks to some of the recent benchmarks on example-based single image SR~\cite{radu_methods_2017,radu_methods_2018,blau_challenge_2018}, several algorithms were introduced for super-resolving images \cite{lim2017enhanced,fan2017balanced,bei2018new,ahn2018image,haris2018deep}. These algorithms can be directly used on multi-spectral images, however, as applied to snapshot mosaic sensors, they do not take into account the spectral correlation of different channels nor the spatial offset of each pixel.


Despite the fact that modern multi-spectral cameras are more adversely affected by resolution constraints than regular RGB cameras, there are not many works specifically on CNN based multi-spectral SR. Example-based learning methods are limited mainly due to the lack of multi-spectral SR benchmarking platforms and difficulty accessing suitable SR spectral datasets. For example,~\cite{li_sr_spectral_2017}, which focuses on hyperspectral SR and not multi-spectral SR, is among one of the few example-based spectral SR methods. The only directly related multi-spectral SR methods~\cite{lahoud_2018,shi_2018}, to the best of our knowledge, were recently introduced through the PIRM2018 spectral SR challenge~\cite{shoeiby_pirm2018_method,shoeiby2018pirm2018}.  The work proposed by Lahoud~\etal~\cite{lahoud_2018}, used an image completion technique followed by 12 convolutional layers to super-resolve images. The second work,~\cite{shi_2018} by Shi~\etal, proposes a deep residual network with channel attention (RCAN) to super-resolve images. The former method~\cite{lahoud_2018}, unlike the latter~\cite{shi_2018} involves some image pre-processing and is not an end-to-end CNN implementation. The RCAN network exploited in~\cite{shi_2018}, has also exhibited state-of-the-art performance in the context of RGB image SR~\cite{zhang_2018}. Both of these works take into account spectral correlation, and do not consider the spatial offsets of each wavelength channel.

While all the above exploit demosaiced (debayered or interpolated) images as their LR/HR pair, a very recent work by Fu \etal \cite{fu_2018} exploits the mosaic RGB images directly to super-resolve hyperspectral images using a variational method. The work was preceded by the SR work by Zhou \etal \cite{zhou_2018} who presented a deep residual network for RGB SR that uses mosaic images. They highlighted the fact that demosaicing, which involves some sort of interpolation (such as bicubic), introduces artifacts that can deteriorate SR performance. 

\section{Proposed Method}
As mentioned earlier, the method presented here is quite general in nature. For the sake of generality we view the problem at hand as that of super-resolving and chromatically mapping images with missing or unevenly distributed wavelength bands. To this end, we propose to investigate the structure of the RCAN network in~\cite{shi_2018} as a baseline, for the combined task of image SR and color-prediction. Moreover, we propose an additional texture network as a means to re-introduce lost information about these bands in the scene. In addition, we notice that the concept of using mosaic images can be extended to multi-spectral images as well, while to the best of our knowledge there is no work reported on using mosaic multi-spectral SR. As a result of this treatment, we can also capitalise on the on-sensor spatial arrangement of the wavelength indexed channels on the mosaic images to improve the SR and colour-prediction performance of our proposed network instead of using demosaiced imagery as input.
\begin{figure*}
  \includegraphics[width=\linewidth]{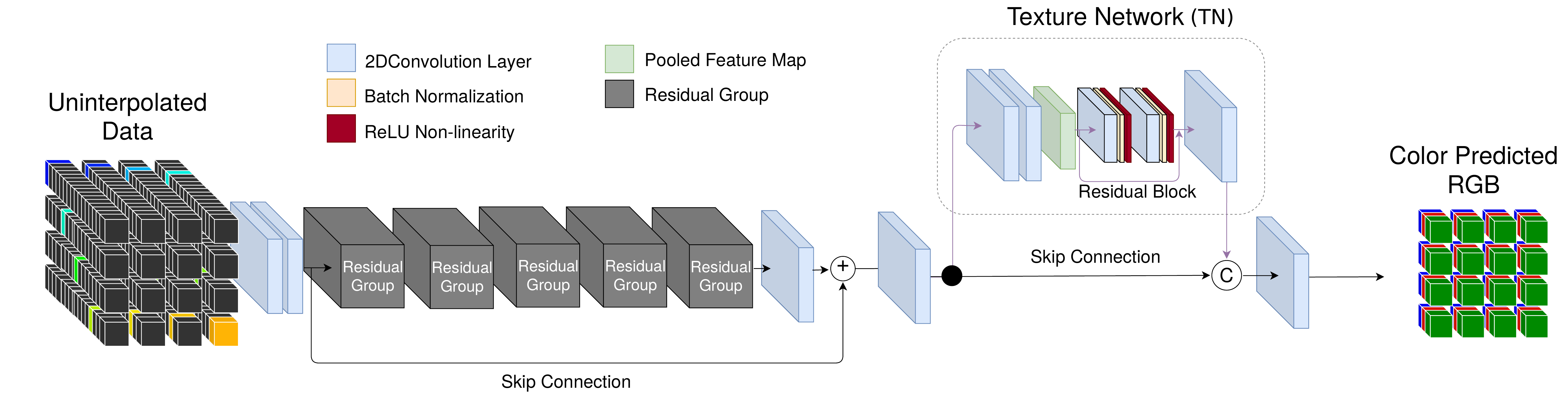}
  \caption{Illustration of the proposed network. The residual group depicted in this figure corresponds to the network of~\cite{shi_2018}. Channel-wise concatenation is denoted by $C$. The input of the network is uninterpolated data (see Figure~\ref{fig:16color}), and the output is the pseudo-colour image.}
  \label{fig:network}
\end{figure*}
\subsection{Texture Sensitive Residual Channel Attention Network (TSRCAN)}

As depicted in Figure~\ref{fig:network}, our network consists of an RCAN network, and a texture network structure. The RCAN network~\cite{zhang_2018,shi_2018} encompasses three main parts, the head, the body, and the tail. The head of the network carries out feature extraction via two convolutional layers. The body is comprised of $g$ number of sequential residual groups as the body of the network. Each residual group contains $b$ number of residual channel attention networks, each constituting a residual block which incorporates within it a channel attention network (CA). The tail of the RCAN network is the reconstruction part which consists of only one convolutional layer to produce an output with the desired dimension of RGB images. The RCAN network, on its own, given the $LR_{MS}$ and $HR_{RGB}$ pairs, can do a relatively modest job to super-resolve and colour-predict the input multi-spectral images. The RCAN network can be shown as
\begin{equation}
    SR^{'}_{RGB} = RCAN(I_{{LR}_{MS}}).
\end{equation}
However, our network expands the RCAN by introducing a texture sensitive network (TN) on the output of the RCAN. Its structure constitutes two convolutional layers  and a pooling layer, followed by a residual block. The residual block comprises two stacks of convolution, batch normalization, and ReLU gating. The whole representation is then upsampled through a deconvolution layer with $k$ number of channels. Each filter in the deconvolution layer represents a texture in the image. The deconvolution layer produces an output of size $k \times w \times h$, where $w$, and $h$ are the width and height of the desired RGB images ($HR_{RGB}$).

The transformation imposed by the TN network is
\begin{equation}
    I_{texture} = TN(SR^{'}_{RGB})
\end{equation}
where $I_{texture}$ has the dimensions $k \times w \times h$. $I_{texture}$ is then concatenated with the output of the RCAN network ($SR^{'}_{RGB}$), producing a tensor of dimension $(k+3) \times w \times h$. The concatenation operation ($\oplus$) is expressed as
\begin{equation}
    I^{'}_{texture} = I_{texture} \oplus SR^{'}_{RGB}.
\end{equation}
Through a convolutional layer, $I^{'}_{texture}$ with $k+3$ channels is reduced to the 3 channels required to produce RGB images. Lets express the operation of this convolutional layer by the transformer $CONV(.)$. The relationship between the input image $I_{{LR}_{MS}}$, and the output $SR_{RGB}$ image can be expressed using the transfer function $TSRCAN(.)$ as
\begin{align}
     SR_{RGB} &=  TSRCAN(I_{{LR}_{MS}}) \nonumber \\
              &= CONV\big( \nonumber \\ 
             TN(RCAN(&I_{{LR}_{MS}}))  \oplus RCAN(I_{{LR}_{MS}})\nonumber \\
            &\big) 
\end{align}

\subsection{Loss functions}
\begin{figure}
  \includegraphics[width=\linewidth]{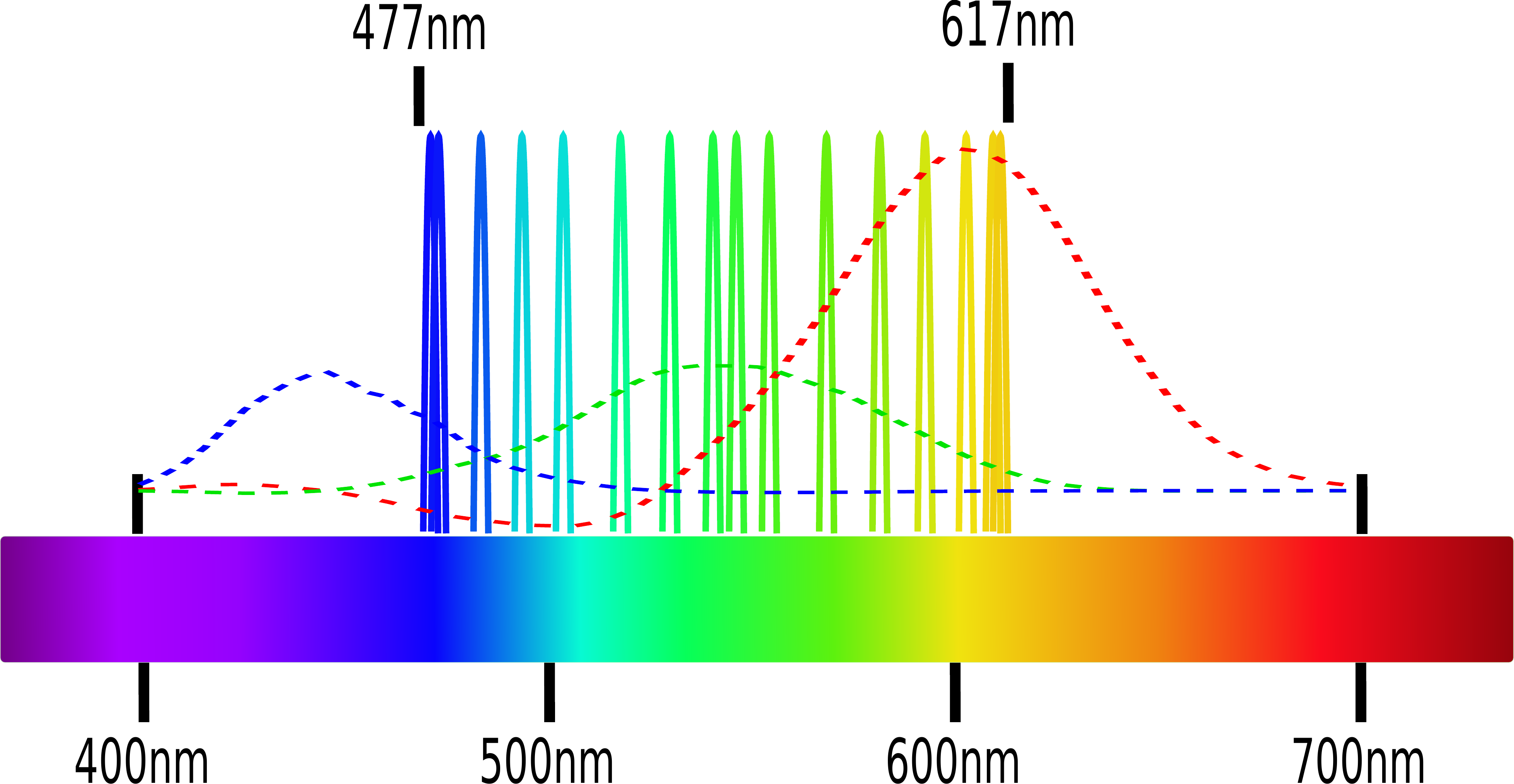}
  \caption{Representation of the visual wavelength against the 16 wavelength bins (sub-pixels) of the multi-spectral camera, and the ideal CMF. The wavelengths for the 16 channels of the multi-spectral camera are $477.2nm$, $478.2nm$,  $489.5nm$,  $500.3nm$,  $510.9nm$,  $523.2nm$, $537.9nm$, $548.9nm$, $553.0$,  $562.5$, $577.3nm$,  $590.5nm$, $599.9nm$,  $612.9nm$, $615.9nm$, $617.5nm$}
  \label{fig:camera_channels}
\end{figure}

In the CNN based SR literature, a simple loss function such as $L_1$\cite{zhang_2018} or $L_2$\cite{shi_2018,lahoud_2018} is usually utilized to train models. An $L_1$ function is less sensitive to outliers compared to an $L_2$ function, and our dataset, which uses registered stereo images, is prone to outliers due to inherent  registration artifacts. Hence, we choose the $SmoothL1$ \cite{smoothl1} PyTorch\cite{pytorch} implementation which is a more stable implementation compared to vanilla $L_1$. For simplicity of notation, we refer to the $SmoothL1$ function as $L_{1s}$, which can be expressed as
\begin{equation}
L_{1s} (\Theta) =\frac{1}{N}\sum_{i=1}^M Z^{i}
\label{eq:sid}
\end{equation}
where
 \[
    Z^{i}=
    \left\{
                \begin{array}{ll}
                  0.5\times (DIF)^2   &if |DIF|<1\\
                  |DIF|-0.5           &otherwise,
                \end{array}
           \right.
  \]
and $DIF =HR^{i}_{RGB}-LR^{i}_{MS}$.  

In addition, we know that (through experiments) RCAN by itself is capable of learning SR and also the colour relationship between input and output images. Hence, the output of RCAN and input of the TN, is also expected to be similar to $I_{{HR}_{RGB}}$. Therefore, we choose to minimise the cost function
\begin{equation}
        L_{1s}^{c} =  L_{1s}^{c}(SR^{'}_{RGB}-I_{{HR}_{RGB}}) + L_{1s}^{c}(SR_{RGB}-I_{{HR}_{RGB}}).
\end{equation}

Figure~\ref{fig:camera_channels} displays the wavelength range of our multi-spectral images vs the wavelength range of the visual light. It also shows the relative amplitude of a CMF for three channels. It is obvious that our multi-spectral images have incomplete blue and red channels, which is one of the main drivers behind this work. We believe that the above loss functions, along with our proposed network, can predict the missing channels and hence improve the colour-prediction performance. 

\subsection{Implementation Details}
Now we specify the implementation details of our proposed TSRCAN. The RCAN part of our network has $g=5$ residual groups. Each residual group contains $b=3$ residual channel attention blocks (RCAB).  The channel attention, similar to \cite{zhang_2018}, has a 64 channel input and 64 weighted channel output with a reduction factor of 16.
The kernel size of all our convolutional layers are set to $3\times3$. Convolutional layers in shallow feature extraction and the body have $c=64$ filters, except at the tail of the RCAN where channels are reduced to $3$. 

The TN structure constitutes a convolutional layer, followed by batch normalization, ReLU, maxpooling, and a residual block similar to that of RCAN but without a channel attention mechanism. In fact, the texture network is identical to the first few layers of the Resnet-18 structure, and we only remove the last layers up to the first residual block. This is followed by a convolutional layer with $k=256$ channels to achieve a tensor with the size $256\times 576 \times 1152$. After concatenating this tensor with $SR^{'}_{RGB}$, the last layer, a convolutional layer with 3 filters produces the desired output dimensions of $3\times 576 \times 1152$.

\subsection{Zero padded, uninterpolated data}
As explained in the introduction, interpolation of mosaic images gives rise to artifacts. For example, SR CNN based methods such as VDSR \cite{kim_vdsr_2016}, and SRCNN \cite{kim_drcnn_2016} that first interpolate the input LR images up to the scale of the HR images suffer from these artifacts via losing information and decreasing computational efficiency \cite{zhang_2018}. Hence, inspired by the procedures in \cite{fu_2018,zhou_2018}, where authors super-resolved hyperspectral \cite{fu_2018}, and RGB images \cite{zhou_2018}, using RGB bayer patterns, we choose not to interpolate the multi-spectral image. Instead, we use the mosaic pattern in the manner presented in Figure \ref{fig:16color}. The mosaic multi-spectral pattern in Figure \ref{fig:16color}(c) represents 1 multi-spectral pixel which constitutes 16 sub-pixels of 16 wavelength channels. To transform the mosaic multi-spectral input to a format that is suitable to be consumed by the network, and to avoid interpolation, we take the following approach. We generate an image with size $16\times 576 \times 1152$, that is a multi-spectral image with height and width of the mosaic image, but with 16 channels. For each channel the value of respective sub-pixel is used and another 15 sub-pixels are added and set to zero. This process, for 1 multi-spectral pixel alone for ease of illustration, is shown in Figure \ref{fig:16color}(c-d).

\section{Experiments}
\subsection{Dataset description}
We carry out our experiments using our 296 registered stereo pair multi-spectral/RGB images which were collected from a diverse range of environments. During acquisition time, no gamma correction was applied to the images. In addition, since the stereo pairs were captured using cameras with different image mosaic sensors, the exposure time was optimised for each camera individually for optimum image quality. One is an RGB camera and the other is a  multi-spectral camera covering the visible wavelength range $~(477-617nm)$. The RGB camera has a CMOS image sensor with a $2\times2$ mosaic (bayer) pattern delivering the three RGB channels whereas the mosaic sensor of the multi-spectral camera has a $4\times4$ pattern delivering 16 wavelength bands. Hence, the resolution of the RGB images in each axis is twice that of the spectral images. Figure \ref{fig:16color} illustrates this resolution relationship between the two filter arrays on both cameras. The original images were interpolated and converted to grayscale for registering using PWC-Net \cite{sun_pwc_2018}, the state-of-the-art optical flow algorithm. Original multi-spectral and the registered RGB were then cropped to the size $576\times1152$ to minimise optical flow artifacts on the border of the images which also led to training acceleration. We split our 296 image pairs to 250 image pairs for training, 25 for validation and 21 for testing. For each image pair, the multi-spectral image with lower spectral and spatial resolution is referred to as $LR_{MS}$ and the registered RGB image with higher spatial resolution is referred to as $HR_{RGB}$.

\subsection{Analysis of the effect of occlusions}
We train our CNN using the $LR_{MS}$ and its registered $HR_{RGB}$ pair. However, with every registration, there are some artifacts including wrong registration and occlusions \cite{wang_flow_occl_2018}. We hypothesise that if these artifacts are abundant, they could affect the training process. To check if errors of the above nature could affect the training process, we take the following approach. We calculate optical flow from the multi-spectral image to the RGB image and vice-versa using the \cite{sun_pwc_2018} algorithm. A straightforward way to detect erroneous flow and occlusions is to calculate the euclidean distance between the two optical flows and remove the pixels with errors larger than a threshold \cite{wang_flow_occl_2018}. Thereby, we removed pixels with errors larger than 3 pixels, and created a mask for each image. We multiplied this mask with $LR_{MS}$, $HR_{RGB}$, and the output of respective model ($SR_{RGB}$). After carrying out several experiments, we did not see a strong correlation between removing the occlusions and improvements in the results. Hence, we believe that either occlusions do not have a significant adverse effect on the training process (which can be attributed to good registration of the images) or substantially more advanced occlusion detection techniques are required to remove their effect. To avoid this topic turning into a research subject of its own, in this work, we decide to acknowledge but not address their effect further in this paper.

\subsection{Settings}
\paragraph{Evaluation metrics:}
 The 21 test images were super-resolved and colour-predicted and then evaluated using Pixel Signal to Noise Ratio (PSNR), Structural Similarity Index (SSIM), and Spectral Information Divergence (SID) per channel. As a reference metric, as is customary with SR evaluations, we also present the results for bicubic up-sampled images where the images were colour-predicted using an ideal CMF function \cite{book_antonio_2012}. We compare the results from our TSRCAN network to the conventional method and the baseline RCAN network.

\paragraph{Training settings:}
 During training, we performed data augmentation on our batches of 10 images of our 250 training images, which included random cropping with size $120\times 120$, random rotation by $0^{\circ}$, $90^{\circ}$ ,$180^{\circ}$ , $270^{\circ}$ with $p=0.25$, and random horizontal flip with $p=0.5$. The batch size is fixed at 10. Our model is trained by ADAM optimizer \cite{adam_2014} with $\beta_1 = 0.9$, $\beta_2=0.999$, and $\epsilon = 10^{-8}$. The initial learning rate is set to $10^{-4}$ and then halved every 2500 epochs. To implement our models we used PyTorch \cite{pytorch}, and in particular, to implement our $L_1$ functions, the $SmoothL_1$ function \cite{smoothl1} was used as the main building block. To test our algorithms, we select the models with the best performance on the validation dataset, and present the test results for those models.
 
\subsection{Baseline RCAN Network vs the conventional method}
Table \ref{table:results} presents the numerical results of our ablation studies. The first two rows compares the performance of the RCAN network (a deep learning approach) vs the traditional method of bicubic upsampling followed by a CMF transfer. There are a number of different CMF available due to the fact that the combination of light wavelengths to produce a given perceived color is not unique \cite{smith_cie_1931}, and there is a degree of subjectivity in drawing a given CMF function \cite{smith_cie_1931}. Hence, to calculate the metrics presented here, only for the conventional method, we normalize all the three R-G-B channels of the ground truth image and the results obtained using the conventional method between 0 and 255 before calculating the metrics. In spite of this, considerable improvements can be observed with the trained baseline RCAN method compared to the conventional method.

\subsection{Effect of using zero padded data}
 \begin{figure}[h]
  \includegraphics[width=\linewidth]{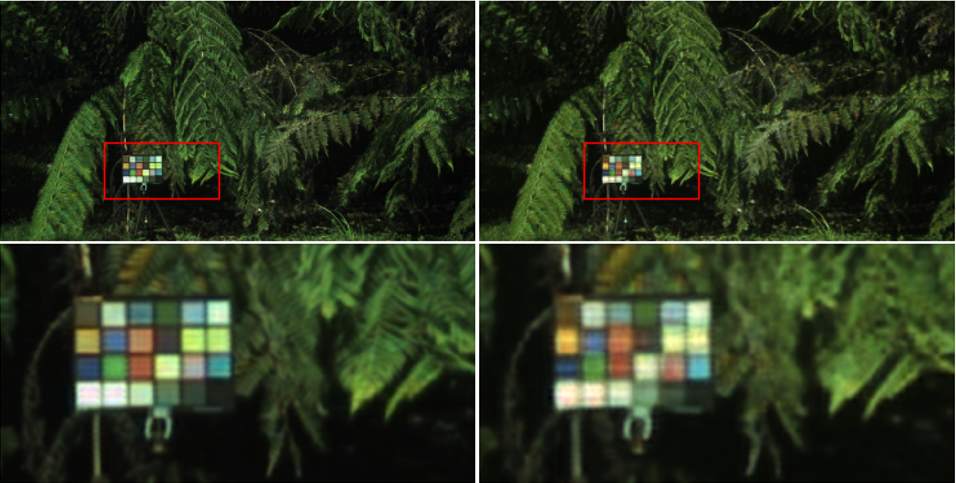}
  \caption{Qualitative comparison of the results for training RCAN with zero-padded uninterpolated data (left) with training RCAN with uninterpolated data without zero-padding \cite{shoeiby_pirm2018_method} (right) which is denoted as RCAN$^*$ in Table \ref{table:results}.}
  \label{fig:rcan_star}
\end{figure}
To assess the effect of zero padded uninterpolated data we need to compare it with a baseline. We chose to train the RCAN network not only with the uninterpolated data, but also with the data format the does not take into account the spatial location of each pixel. In this way, the multi-spectral mosaic pixel in Figure \ref{fig:16color}(c) with dimension $4\times 4 \times 16$ translates to $1\times 1 \times 16$. In other words, we remove the zero padded sub-pixels in Figure \ref{fig:16color}(d). In fact, this is the approach that was taken in the algorithms introduced in \cite{shoeiby_pirm2018_method}. In Table \ref{table:results}, the results obtained using this method is presented as RCAN$^*$. Comparing the results of RCAN with RCAN$^*$, it can be seen that with zero-padded uninterpolated data, while displaying an improvement across all metrics, the improvements in PSNR is minimal. However, looking at Figure \ref{fig:rcan_star}, it is obvious that the quality of the images produced by RCAN are superior to that of RCAN$^*$. We can reaffirm the notion that, while PSNR remain a descent measure of image quality, it does not provide an accurate measure of perceptual quality of the image \cite{mse_ssim_wang_2009} compared to SSIM and SID \cite{sid_chang_1999}. 

\begin{table}
\scriptsize	
\begin{center}
\begin{tabular}{|l|c|c|c|c|c|}
\hline
\bf{Method}&\bf{PSNR} &\bf{SSIM}  &\multicolumn{3}{c|}{\bf{SID}} \\\cline{4-6}
             & \bf{(dB)} &        & \bf{Blue} & \bf{Green}& \bf{Red} \\ \hline
Bicubic +    &   22.375   & 0.779  &  7.29e-05 & 5.72e-05  & 7.43e-05 \\ 
    CMF      &   (3.40)  & (0.102)& (3.39e-4) &(1.79e-4)  &(1.96e-4)    \\ \hline
RCAN$^{*}$   &   24.78   &  0.814 & 5.53e-05 & 4.18e-05&  4.38e-05   \\ 
             &  (4.62)   & (0.093)& (4.38e-05)&(3.34e-05) & (3.90e-05) \\ \hline
RCAN         &   24.90   & 0.847  &\textbf{4.90e-05}&\textbf{3.63e-05}& 4.147e-05 \\ 
 (Ours)      &  (3.50)  & (0.0812)&(3.64e-05)&(3.25e-05) & (4.38e-05)   \\ \hline
TSRCAN    &  \textbf{26.02}& \textbf{0.855}  & 5.74e-05 &3.99e-05& \textbf{3.53e-05}\\ 
(Ours)    &  (4.59)      & (0.095) & (6.29e-05)& (3.83e-05)&  (2.81e-05)  \\ \hline
\end{tabular}
\end{center}
\caption{Mean and standard deviation (in parenthesis) of PSNR, SSIM, and SID obtained using different models. RCAN$^{*}$ denotes RCAN trained with data without zero padding.}
\label{table:results}
\end{table}

 \begin{figure*}[h]
  \includegraphics[ width=\linewidth]{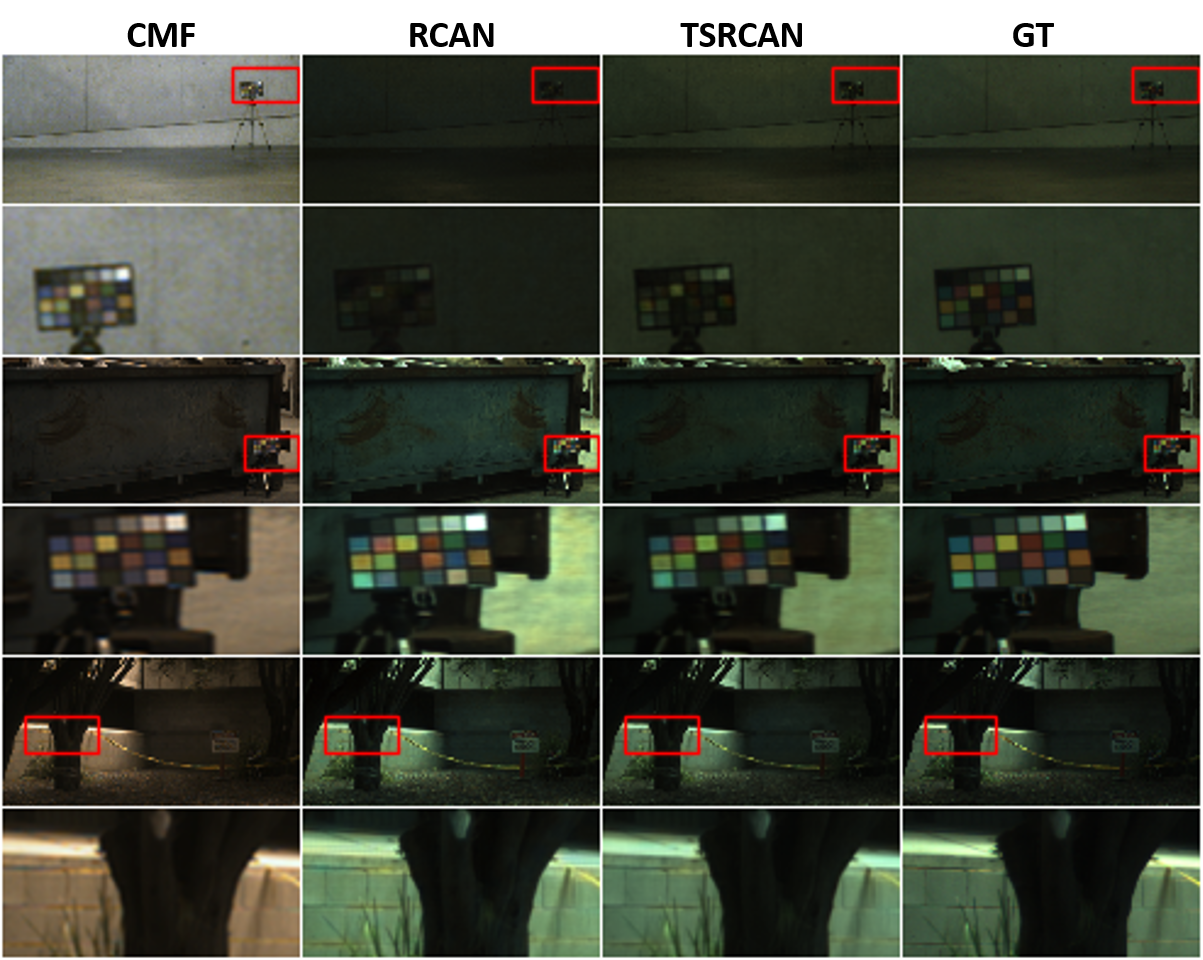}
  \caption{Visual representation of the performance of three methods: the conventional method (white-balanced), RCAN, and TSRCAN along with the ground truth with diverse lighting environments. For better visualization, we show a zoomed in area of the images in the even rows.}
  \label{fig:images}
\end{figure*}

 \begin{figure}[h]
  \includegraphics[width=\linewidth]{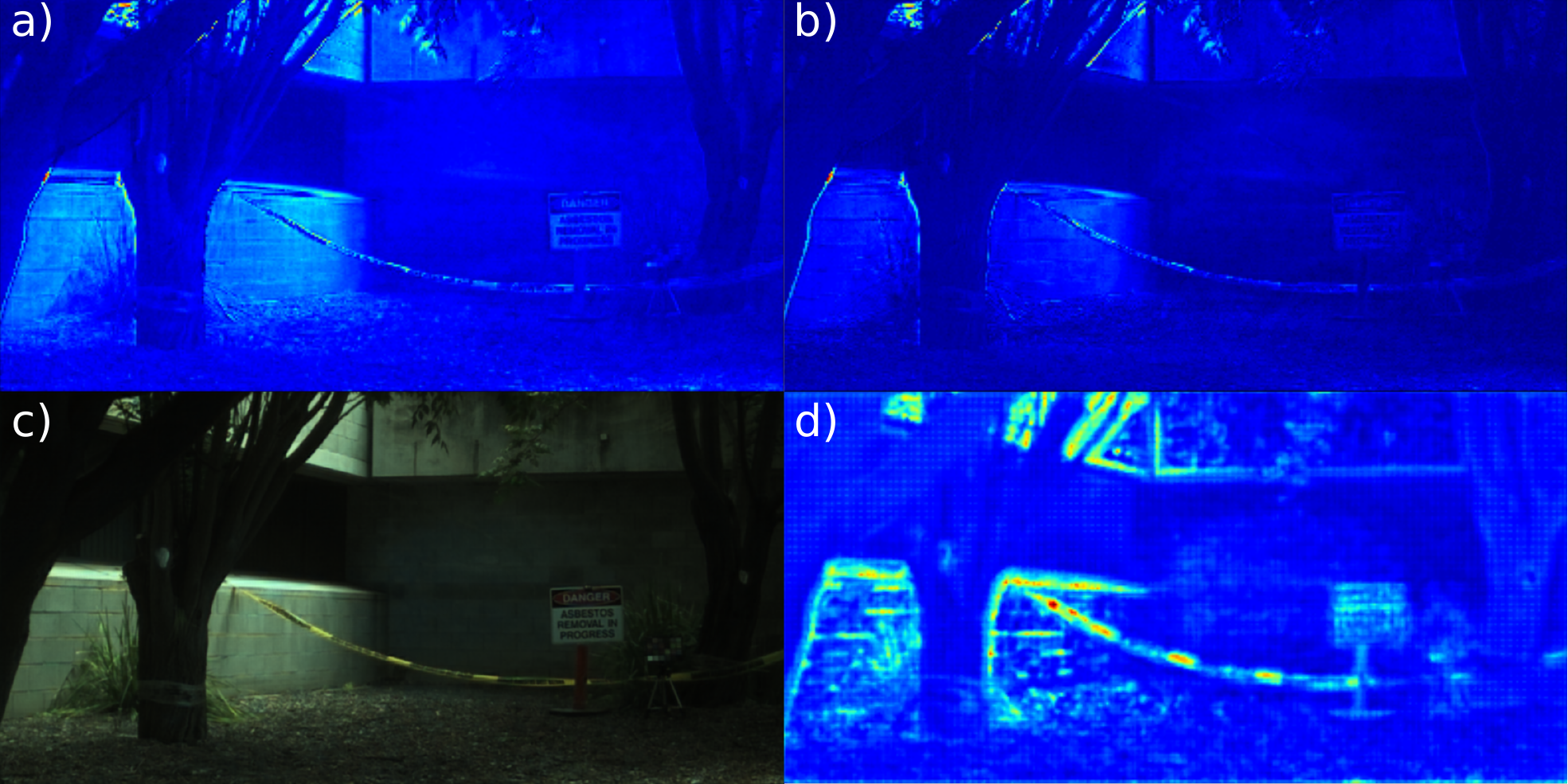}
  \caption{Effect of the TN. For an image from the test set (a) shows the error map at the output of RCAN, (b) shows the error map at the output of TSRCAN, (c) depicts the output image from TSRCAN, and (d) illustrates the activation map at the output of TN.}
  \label{fig:TN}
\end{figure}
\subsection{Effect of Texture Network (TN)}
We trained TSRCAN from scratch. A considerable improvement of $1.1 dB$ in PSNR performance is observed compared to the baseline RCAN network performance. SSIM also displays some improvements. Regarding the  SID metric, there are some improvements in the red channel compared to the baseline RCAN network but not in the green or the blue channels. Also, the qualitative results in Figure \ref{fig:images} evidence the superiority of our network relative to the baselines. To explain the overall improvement, we look into the activation map at the output of the TN. Figure \ref{fig:TN} presents, for an image from the test set, (a) the error map at the output of RCAN, (b) the error map at the output of TSRCAN, (c) output image of TSRCAN, and (d) the activation map at the output of the TN. It seems generally, in textures/pixels that RCAN produces a larger error, the activation map is more ``active", leading to less error for TSRCAN in the same areas/pixels.
\subsection{Effect of network size}

\begin{figure}
  \includegraphics[width=\linewidth]{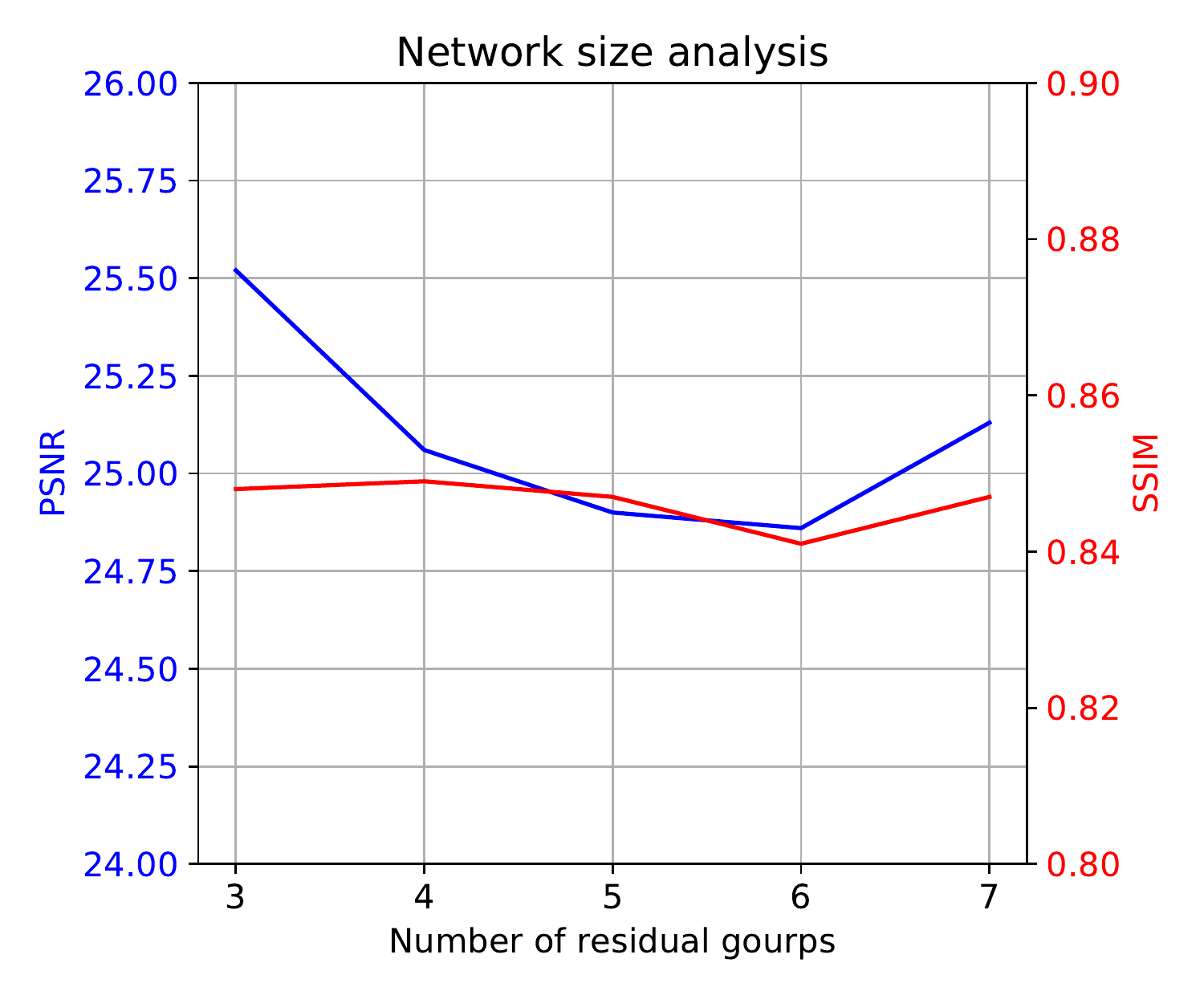}
  \caption{Effect of the number of residual groups on the performance of the network in terms of PSNR and SSIM on the test data.}
  \label{fig:network_size}
\end{figure}

Figure \ref{fig:network_size} displays PSNR and SSIM results for RCAN networks with different numbers of residual groups. There is no clear correlation between the number of residual groups and the performance of the network. For example, as the number of residual groups increases from 5 to 6 and 7, the PSNR improves only negligibly by $0.27dB$. Also, SSIM results deteriorate slightly and then bounce back slightly. For 4, and 3 residual groups, the improvement in PSNR is 0.16dB, and 0.62dB respectively. In addition, these results do not correlate with the SSIM results in which the baseline we use (with 5 residual groups) exhibits approximately the mean of all the RCAN versions. Our choice of using five residual groups was due to the fact that the most recent work on multi-spectral SR, based on RCAN, used 5 residual groups, and we chose to investigate that model as our baseline \cite{shi_2018}. These observations, indeed, highlight the effectiveness of the texture network module in our network (TSRCAN) which exhibited a 1.1dB improvement in PSNR compared to the baseline RCAN. 

\subsection{Spectral Information Divergence (SID)}
A close observation of SID per channel results in Table \ref{table:results} shows that the blue channel generally exhibits a larger SID error, followed by the red channel, and with the green channel having the lowest SID. This is true except for the results obtained using the conventional method, where we white-balanced the channels to calculate the metrics. These results correlate with the fact that the multi-spectral camera has a large portion of the blue wavelength range missing (see Figure \ref{fig:camera_channels}) compared to the red wavelengths. The green channel exhibits the lowest SID because the multi-spectral camera covers the whole green wavelength range, although sparsely. The reddish appearance in the white-point balanced images transferred using the conventional method can also be explained by the fact that the camera has a larger portion of the blue wavelengths missing compared with the blue wavelengths. Therefore, leading to an exaggerated contribution of the red channel compared to the blue channel.

\subsection{Effect of poor lighting conditions}
In Figure \ref{fig:images}, we include scenarios with diverse lighting conditions. Specifically, the first row of the figure includes an indoor image with poor lighting which implies an incomplete wavelength spectrum. To elaborate, the camera channels depicted in Figure \ref{fig:camera_channels} can be thought of as sampling wavelengths which already do not cover the complete visible wavelength range. On top of this, poor lighting conditions results in fewer efficient samples. For example, with the indoor image in the first row, most of the wavelength spectrum is likely to be emitted from fluorescent lights (which is known for producing a poor, nonuniform, sparse spectrum) and some leakage of outdoor light which results in a poor spectrum. Hence, there are much fewer wavelength samples and the network is having a harder time predicting the unknown wavelengths, and hence the R-G-B values of multi-spectral pixels. Our algorithm produces somewhat decent results for this difficult scenario, which means that the network is doing a good job in learning to predict the unknown wavelengths. However, the performance can be improved by expanding the dataset with images taken in a controlled laboratory environment to include more examples of poorly lit conditions. Given that the dataset contains 296 image pairs to train and test our networks, expanding this data set is a perfectly feasible task. In fact, this is a future work that we are planning to carry out.

\section{Conclusion}
In this paper we proposed a novel deep learning approach that addresses the ill-posed problem of producing RGB images from spectrally and spatially under-sampled multi-spectral images, which significantly outperforms, quantitatively and qualitatively, conventional methods as well as the state-of-the-art RCAN network. Moreover, the method is quite general in nature and can be applied to multi-spectral images of low spatial and spectral resolution with unevenly spaced or missing channels. Our approach uses a texture sensitive block to enable the network so as to re-introduce information from missing wavelength bands that may be still implicitly available in the texture of the image. In addition, we have introduced a novel dataset consisting of 296 registered stereo multi-spectral/RGB image pairs. 

{\small
\bibliographystyle{ieee}
\bibliography{references}
}

\end{document}